\documentclass[a4paper,11pt]{article}
\usepackage[english]{babel}
\usepackage{graphicx,color}
\usepackage{ifpdf}
\ifpdf
  \usepackage[pdftex]{hyperref}
\else
  \usepackage[hypertex]{hyperref}
\fi
\hypersetup{bookmarks=true, unicode=true, colorlinks=true, linkcolor=black, citecolor=black, filecolor=black, urlcolor=black}
\usepackage{amsmath,amssymb,amsfonts,bm,euscript,amscd,ae,amsthm,exscale,amsopn,amstext}
\usepackage{ifthen}
\usepackage[T1]{fontenc}
\usepackage[cp1251]{inputenc}
\usepackage[active]{srcltx}

\hypersetup{colorlinks,%
linkcolor=blue,%
urlcolor=blue,%
citecolor=red,%
filecolor=magenta}

\title{Angularly Deformed Special Relativity and Its Results for Quantum Mechanics}

\author{\L ukasz Andrzej Glinka}

\begin{document}

\date{\today}

\maketitle

\begin{abstract}
In this paper, the deformed Special Relativity, which leads to an essentially new theoretical context of quantum mechanics, is presented. The formulation of the theory arises from a straightforward analogy with the Special Relativity, but its foundations are laid through the hypothesis on breakdown of the velocity-momentum parallelism which affects onto the Einstein equivalence principle between mass and energy of a relativistic particle. Furthermore, the derivation is based on the technique of an eikonal equation whose well-confirmed physical role lays the foundations of both optics and quantum mechanics. As a result, we receive the angular deformation of Special Relativity which clearly depicts the new deformation-based theoretical foundations of physics, and, moreover, offers both constructive and consistent phenomenological discussion of the theoretical issues such like imaginary mass and formal superluminal motion predicted in Special Relativity for this case. In the context of the relativistic theory, presence of deformation does not break the Poincar\'{e} invariance, in particular the Lorentz symmetry, and provides essential modifications of both bosons described through the Klein-Gordon equation and fermions satisfying the Dirac equation. On the other hand, on the level of discussion of quantum theory, there arises the concept of emergent deformed space-time, wherein the presence of angular deformation elucidates a certain new insight into the nature of spin, as well as both the Heisenberg uncertainty principle and the Schr\"odinger wave equation.
\end{abstract}
\newpage
\section{Introduction}

In Special Relativity, Cf. the Ref. \cite{sr}, the Einstein equivalence principle
\begin{equation}
E^2=m^2c^4+p^2c^2,\label{Ein0}
\end{equation}
where $c$ is the speed of light in vacuum, relates kinetic energy $E$ to magnitude $p=\sqrt{p^ip_i}$ of a momentum vector $p_i$ of a particle equipped with a mass $m$. In general, a velocity vector $v_i$ and speed $v$ of a particle are determined as
\begin{equation}
v^i=\frac{\partial E}{\partial p_i}, \quad v^ip_i=vp,\quad v=(v^iv_i)^{1/2}=\frac{\partial E}{\partial p}=\frac{pc^2}{E}.
\end{equation}
In particular, for a massive particle one has $E=\gamma mc^2$, $p=\sqrt{\gamma^2-1}mc$, and $v=p/(\gamma m)$ where $\gamma=\left(1-\frac{v^2}{c^2}\right)^{-1/2}$ is the Lorentz factor, for a massless particle $E=pc$ and $v=c$, for an imaginary mass one has to deal with superluminal particles known as tachyons. The space-time coordinates four-vectors are either contravariant $x^\mu=[ct,x^i]$ or covariant $x_{\mu}=\eta_{\mu\nu}x^{\nu}$, and the energy-momentum four-vectors are $p^\mu=[E,p^ic]$ or $p_{\mu}=\eta_{\mu\nu}p^{\nu}$, both space-time and energy-momentum space are equipped with the Minkowski metric $\eta_{\mu\nu}=\mathrm{diag}[1,-1,-1,-1]$.

Locally, the energy-momentum interval $ds^2=\eta_{\mu\nu}dp^{\mu}dp^\nu=dE^2-c^2 dp_idp^i$ is symmetric, that is $ds^2=ds'^2$, with respect to the action of the Poincar\'{e} transformations $p'^\mu=\Lambda^\mu_{~~\!\nu}p^\nu+p_0^\mu$ with a constant energy-momentum four-vector $p_0^\mu$, while the space-time interval $ds^2=\eta_{\mu\nu}dx^{\mu}dx^\nu=c^2dt^2-dx^idx_i$ displays invariance with respect to the action of the Poincar\'{e} transformations $x'^\mu=\Lambda^\mu_{~~\!\nu}x^\nu+x_0^\mu$ with a constant space-time four-vector $x_0^\mu$, whenever the Lorentz matrices obey $(\Lambda^{-1})^\mu_{~~\!\kappa}=\eta^{\mu\lambda}\Lambda_\lambda^{~~\!\nu}\eta_{\nu\kappa}$ and $\det\Lambda^\mu_{~~\!\nu}=\pm 1$. Relativistic invariance includes symmetry with respect to action of the Lorentz transformations $p'^\mu=\Lambda^\mu_{~~\!\nu}p^\nu$ and $x'^\mu=\Lambda^\mu_{~~\!\nu}x^\nu$. Globally, the Lorentz symmetry holds for either $s^2=\eta_{\mu\nu}p^{\mu}p^\nu=E^2-c^2p^2$ or $s^2=\eta_{\mu\nu}x^{\mu}x^\nu=c^2t^2-x^2$, that is $s^2=s'^2$, whereas preservation of the Poincar\'{e} symmetry demands
\begin{eqnarray}
\eta_{\mu\nu}\left(p_0^\mu p_0^\nu+\left\{\Lambda^\mu_{~~\!\kappa} p^\kappa,p_0^\nu\right\}\right)&=&0,\\
\eta_{\mu\nu}\left(x_0^\mu x_0^\nu+\left\{\Lambda^\mu_{~~\!\kappa}x^\kappa,x_0^\nu\right\}\right)&=&0,
\end{eqnarray}
for every $x^\mu$ and $p^\mu$. For a particle motion, the well-known representation is
\begin{equation}
\Lambda^\mu_{~~\!\nu}=\left[\begin{array}{cc}\gamma&-\gamma R^i_j\frac{v^i}{c}\\
-\gamma\frac{v_j}{c}&R^i_k\left(\delta^k_j+(\gamma-1)\frac{v^kv_j}{v^2}\right)\end{array}\right],
\end{equation}
where a matrix of rotation $R^i_j\in \mathsf{SO}(3)$ in three-dimensional Euclidean space $\mathbf{R}^3$ with metric $\delta^i_j=\mathrm{diag}[1,1,1]$, can be defined through either the Euler angles $(\phi,\varphi,\theta)$
\begin{equation}
  R^i_j(\theta,\varphi,\phi)=\exp\left[(r^i_j)^{(3)}\phi\right]\exp\left[(r^i_j)^{(2)}\varphi\right]\exp\left[(r^i_j)^{(3)}\theta\right],
\end{equation}
or the Tait-Bryan angles $(\phi,\varphi,\theta)$
\begin{equation}
  R^i_j(\theta,\varphi,\phi)=\exp\left[(r^i_j)^{(3)}\theta\right]\exp\left[(r^i_j)^{(2)}\varphi\right]\exp\left[(r^i_j)^{(1)}\phi\right],
\end{equation}
with the Rodriques formula
\begin{equation}
\exp\left[(r^i_j)^{(p)}\alpha\right]=\delta^i_j+(r^i_j)^{(p)}\sin\alpha+(r^i_k)^{(p)}(r^k_j)^{(p)}(1-\cos\alpha),
\end{equation}
where $(r^i_j)^{(p)}$ are infinitesimal rotations matrices around a $p$-axis
\begin{equation}
(r^i_j)^{(1)}=\left[\begin{array}{ccc}0&0&0\\0&0&-1\\0&1&0\end{array}\right], (r^i_j)^{(2)}=\left[\begin{array}{ccc}0&0&1\\0&0&0\\-1&0&0\end{array}\right],
(r^i_j)^{(3)}=\left[\begin{array}{ccc}0&-1&0\\1&0&0\\0&0&0\end{array}\right],
\end{equation}
and form the Lie algebra $\mathfrak{so}(3)=\mathrm{span}\left\{(r^i_j)^{(1)},(r^i_j)^{(2)},(r^i_j)^{(3)}\right\}$ of $\mathsf{SO}(3)$ group
\begin{equation}
[(r^i_k)^{(a_1)},(r^k_j)^{(a_2)}]=\varepsilon^{a_1a_2a_3}(r^i_j)^{(a_3)},\quad \varepsilon^{a_1a_2a_3}=\prod_{1\leq i<j\leq3}\mathrm{sgn}(a_j-a_i),
\end{equation}
where $\varepsilon^{pq}_r$ is a three-dimensional Levi-Civita symbol, whose Casimir operator is equal to the unit matrix $\delta^\mu_\nu=\mathrm{diag}[1,1,1,1]$. For lack of spatial rotations, one has to deal with the Lorentz boosts defined through $R^i_j=\delta^i_j$.

From the point of view of modern theoretical physics, the particularly important ramifications of the Special Relativity based on the Einstein equivalence principle (\ref{Ein0}) are quantum mechanics, Cf. the Refs. \cite{qm}, which throughout the theory of bosons based on the Klein--Gordon equation
\begin{equation}
\left(\eta_{\mu\nu}{\partial}^\mu{\partial}^\nu+\frac{{m}^2{c}^2}{{\hbar}^2}\right){\Psi}({x},{t})=0,\label{KG}\\
\end{equation}
and fermions based on the Dirac equation
\begin{equation}
\left(i{\hbar}{c}\gamma^\mu{\partial}_\mu-{m}{c}^2\right){\psi}({x},{t})=0,\label{Dir}
\end{equation}
for either massive or massless particles, where $\gamma^\mu$ are the standard Dirac gamma matrices obeying the relations $\left\{\gamma^\mu,\gamma^\nu\right\}=2\eta^{\mu\nu}$ of the Clifford algebra $\mathcal{C\ell}_{1,3}(\mathbf{C})$, as well as, in the non-relativistic limit, the Schr\"odinger equation
\begin{equation}
 i\hbar\frac{\partial}{\partial t}{\Phi}(x,t)=\left(-\frac{\hbar^2}{2m}\nabla^2+{\mathrm{V}}(x)\right){\Phi}(x,t),\label{Schr}
\end{equation}
which, along with both the Heisenberg uncertainty principle
\begin{equation}
\left[\hat{{x}}_i,\hat{{p}}_j\right]=i\hbar\delta_{ij},
\end{equation}
and the Born normalization condition
\begin{equation}
  \int d^3x \left|\Phi(x,t)\right|^2=1,
\end{equation}
lays the foundations of the modern-day understanding of physics at the microscopic scales and, moreover, throughout the methods of quantum field theory, has already became the theoretical nucleus of the Standard Model of elementary particles and fundamental interactions which deals with physics at the high energy scales, Cf. the Refs. \cite{hep}. It should be noticed that equally important ramification of Special Relativity is its generalization to a curved space-time known as General Relativity, which is the well-accepted classical theory of gravitation and lays the foundations of modern both cosmology and astrophysics, Cf. the Refs. \cite{gr}, at the astronomical and ultra-high energy scales, but in this paper we shall omit discussion of this particular context of Special Relativity. Nevertheless, in general, presence of a small non-trivial change in the formalism of Special Relativity will have been resulting in inevitable modifications to the standard points of view onto the physical world on both microscopic and macroscopic levels of discussion. In particular, theoretical ramifications will have been touching analysis of experimental and observational data from both particle accelerators which explore the microscopic world as well as telescopes and cosmic missions which collect information on phenomena occurring at the astronomical scales and the large scale structure limits.

In the present article, we provide a step-by-step derivation to the very specific model of deformed Special Relativity, which in our opinion is caused through a physical realization of the hypothesis on breakdown of the velocity-momentum parallelism and, for this reason, immediately impacts onto the Einstein equivalence principle between mass, momentum, and energy of a particle. Our discussion shows that acceptance of this point of view onto the physical reality leads to the concept of emergent deformed space-time and, moreover, the substantial modifications in the quantum mechanical understanding of the nature of a particle. In this manner, the variant of deformed Special Relativity, which we shall present here under the name angularly deformed Special Relativity, is able to lay the foundations for the new physics, particularly the high energy physics beyond the Standard Model.

\section{Angularly Deformed Special Relativity}

\subsection{Violated Velocity-Momentum Parallelism}

Let us consider a Lorentz-invariant deformed energy
\begin{equation}
\tilde{E}^2=E^2+\mathcal{E}^2(\tilde{E}^2,p^2)\label{constra}
\end{equation}
with a non-deformed momentum vector $\tilde{p}_i=p_i$ , and violated velocity-momentum parallelism
\begin{equation}
\tilde{v}^i=\frac{\partial \tilde{E}}{\partial p_i},\quad \tilde{v}^ip_i=\tilde{v}p\cos\delta,\quad \tilde{v}=(\tilde{v}^i\tilde{v}_i)^{1/2}=\frac{1}{\cos\delta}\frac{\partial \tilde{E}}{\partial p}=\frac{p\tilde{c}^2}{\tilde{E}}.
\end{equation}
Making use of the method of implicit differentiation one can write
\begin{equation}
F=E^2+\mathcal{E}^2-\tilde{E}^2=0,\quad F_p=\partial_p F,\quad F_{\tilde{E}}=\partial_{\tilde{E}} F,\quad \partial_p\tilde{E}=-\frac{F_p}{F_{\tilde{E}}},\label{constra1}
\end{equation}
and establish the formulas
\begin{eqnarray}
\tilde{c}&=&c\left(\frac{1}{\cos\delta}\frac{1+\frac{1}{2pc^2}\partial_{p}\mathcal{E}^2}{1-\frac{1}{2\tilde{E}}\partial_{\tilde{E}}\mathcal{E}^2}\right)^{1/2},\label{si}\\
\tilde{v}&=&v\frac{E}{\tilde{E}}\frac{\tilde{c}^2}{c^2}.\label{velo}
\end{eqnarray}
If the Eq. (\ref{constra}) is rewritten in the form of the Einstein equivalence principle on the deformed quantities
\begin{equation}
\tilde{E}^2=p^2\tilde{c}^2+\tilde{m}^2\tilde{c}^4,\label{bomb1}
\end{equation}
and the rest energy is invariant with respect to deformation
\begin{equation}
\tilde{m}\tilde{c}^2=mc^2,
\end{equation}
then the relativistic invariance holds on both global and local levels
\begin{eqnarray}
\tilde{s}^2=\eta_{\mu\nu}\tilde{p}^\mu\tilde{p}^\nu=\tilde{E}^2-p^2\tilde{c}^2=\tilde{m}^2\tilde{c}^4=m^2c^4=\eta_{\mu\nu}p^\mu p^\nu=s^2=s'^2,\label{interval}\\
d\tilde{s}^2=\eta_{\mu\nu}d\tilde{p}^{\mu}d\tilde{p}^\nu=d\tilde{E}^2-\tilde{c}^2 dp^2=\eta_{\mu\nu}dp^\mu dp^\nu=ds^2=ds'^2,\label{intervalA}
\end{eqnarray}
where prime means a relativistically transformed quantity, and, consequently,
\begin{eqnarray}
\tilde{m}&=&m\frac{c^2}{\tilde{c}^2}=m\frac{p^2c^2}{\tilde{E}^2-m^2c^4},\label{si1}\\
\mu&=&\frac{\tilde{m}^2}{m^2}-1=\dfrac{c^4}{\tilde{c}^4}-1,\\
E^2&=&\frac{\tilde{m}}{m}p^2\tilde{c}^2+\tilde{m}^2\tilde{c}^4=(1+\mu)^{1/2}p^2\tilde{c}^2+m^2c^4,\label{bomb}\\
\tilde{c}&=&c\left(1+\mu\right)^{-1/4}=\frac{1}{p}\left(\tilde{E}^2-m^2c^4\right)^{1/2},\label{pyza}\\
\mathcal{E}^2&=&\left(1-\frac{\tilde{m}}{m}\right)p^2\tilde{c}^2=\left(\frac{\tilde{c}^2}{c^2}-1\right)p^2c^2,\label{defor}\\
\tilde{v}&=&\frac{1}{p\tilde{E}}(\tilde{E}^2-m^2c^4),\label{velocyA}
\end{eqnarray}
where $\mu$ is the relative squared-mass difference. Furthermore, consistency gives the eikonal-type first order partial differential equation
\begin{equation}
   (pc)^2\left(1+\frac{1}{2pc}\frac{\partial\mathcal{E}^2}{\partial (pc)}\right)=(\tilde{E}^2-m^2c^4)\left(1-\frac{1}{2\tilde{E}}\frac{\partial\mathcal{E}^2}{\partial \tilde{E}}\right)\cos\delta,\label{bozena}
\end{equation}
which through separation of variables $\mathcal{E}^2=\mathcal{E}^2_1(\tilde{E})+\mathcal{E}^2_2(p)$ leads to the system
\begin{eqnarray}
\frac{d\mathcal{E}^2_1}{d \tilde{E}}&=&2\tilde{E}-\frac{\epsilon^2}{\cos\delta}\frac{\tilde{E}}{\tilde{E}^2-m^2c^4},\\ \frac{d\mathcal{E}^2_2}{d(pc)}&=&\frac{\epsilon^2}{pc}-2pc,
\end{eqnarray}
where $\epsilon^2$ is a real non-zero separation constant, which along with the constraint for the initial data $\tilde{E}_0^2=p_0^2c^2+m^2c^4+\mathcal{E}^2_0$ gives
\begin{equation}
  \mathcal{E}^2=\tilde{E}^2-\tilde{E}_0^2-(p^2-p_0^2)c^2-\frac{\epsilon^2}{2\cos\delta}\ln\left|\frac{\tilde{E}^2-m^2c^4}{\tilde{E}_0^2-m^2c^4}\right|+\frac{\epsilon^2}{2}\ln\frac{p^2}{p_0^2},\label{defka}
\end{equation}
whereas, through the relation (\ref{constra}) applied into the solution (\ref{defka}), one obtains
\begin{eqnarray}
\tilde{E}^2&=&m^2c^4+(\mathcal{E}^2_0+p_0^2c^2)\left(\frac{p}{p_0}\right)^{2\cos\delta}\exp\left(-\frac{2\mathcal{E}^2_0}{\epsilon^2}\cos\delta\right),\\
\mathcal{E}^2&=&-p^2c^2+(\mathcal{E}^2_0+p_0^2c^2)\left(\frac{p}{p_0}\right)^{2\cos\delta}\exp\left(-\frac{2\mathcal{E}^2_0}{\epsilon^2}\cos\delta\right),\\
0&=&(\mathcal{E}^2_0+p_0^2c^2)\left[1-\exp\left(-\frac{2\mathcal{E}^2_0}{\epsilon^2}\cos\delta\right)\right].\label{cond}
\end{eqnarray}
For non-zero $\cos\delta$, the Eq. (\ref{cond}) gives either $\mathcal{E}^2_0=-p_0^2c^2$, that is the unchanged rest energy $\tilde{E}=mc^2$ for which deformation is $\mathcal{E}^2=-p^2c^2$, or $\mathcal{E}^2_0=0$ and
\begin{eqnarray}
E_0^2&=&p_0^2c^2+m^2c^4,\label{eini}\\
\tilde{E}^2&=&m^2c^4+p_0^2c^2\left(\frac{p}{p_0}\right)^{2\cos\delta},\label{equiv}\\
\mathcal{E}^2&=&-p^2c^2+p_0^2c^2\left(\frac{p}{p_0}\right)^{2\cos\delta}.\label{delto}
\end{eqnarray}

\subsection{Angularly-Deformed Relativistic Particles}

Therefore, for a massless particle, $m=0$, one obtains
\begin{eqnarray}
\tilde{m}&=&0,\\
\tilde{c}&=&c\left(\frac{p}{p_0}\right)^{\cos\delta-1},\label{siA1ab}\\
\tilde{v}&=&\tilde{c},\label{pop}\\
\tilde{E}&=&p_0c\left(\frac{\tilde{v}}{c}\right)^{\cos\delta/(\cos\delta-1)},\\
\mathcal{E}^2&=&p_0^2c^2\left[\left(\frac{\tilde{v}}{c}\right)^{2\cos\delta/(\cos\delta-1)}-\left(\frac{\tilde{v}}{c}\right)^{2/(\cos\delta-1)}\right],\\
\tilde{L}&=&-p_0c(1-\cos\delta)\left(\frac{\tilde{v}}{c}\right)^{\cos\delta/(\cos\delta-1)},\label{lag2}
\end{eqnarray}
where $\tilde{L}=p_i\tilde{v}^i-\tilde{E}$ is the deformed kinetic Lagrangian.

For a massive particle, the deformed Lorentz factor $\tilde{\Gamma}$ convention gives
\begin{eqnarray}
\tilde{E}&=&\tilde{\Gamma} mc^2,\label{ene}\\
\mathcal{E}^2&=&-p^2c^2+(\tilde{\Gamma}^2-1)m^2c^4,\\
\tilde{\Gamma}&=&\left(1+\left(\frac{p}{p_0}\right)^{2\cos\delta}\frac{p_0^2}{m^2c^2}\right)^{1/2},\\
  \tilde{c}&=&c\left(\frac{p}{p_0}\right)^{\cos\delta-1},\label{siA1}\\
  \tilde{m}&=&m\left(\frac{p}{p_0}\right)^{2(1-\cos\delta)},\label{si1A}\\
  \mu&=&\left(\frac{p}{p_0}\right)^{4(1-\cos\delta)}-1,\\
   \tilde{v}&=&c\left(\frac{p}{p_0}\right)^{2\cos\delta-1}\left(\frac{m^2c^2}{p_0^2}+\left(\frac{p}{p_0}\right)^{2\cos\delta}\right)^{-1/2}=\frac{\tilde{\Gamma}^2-1}{\tilde{\Gamma}}\frac{mc^2}{p},\label{velocyA1}\\
\tilde{L}&=&-\frac{mc^2}{\tilde{\Gamma}}\left(\tilde{\Gamma}^2+(1-\tilde{\Gamma}^2)\cos\delta\right).\label{lag1}
\end{eqnarray}

For a Lagrangian $L=L(\tilde{v})$, in particular both (\ref{lag2}) and (\ref{lag1}), one has
\begin{eqnarray}
\frac{\partial L}{\partial \tilde{v}^i}&=&p_i=\frac{\tilde{v}_i}{\tilde{v}}p\cos\delta,\quad p=\frac{1}{\cos\delta}\frac{\partial L}{\partial \tilde{v}},\label{momma}\\
\frac{d p_i}{dt}&=&0.\label{el}
\end{eqnarray}
The Eq. (\ref{momma}) agrees with (\ref{siA1ab}) and (\ref{velocyA1}). The Eqs. (\ref{el}) give $p_i=p_{0i}$, or
\begin{equation}
  \frac{p_0}{p}=\cos\delta\in(0,1],\quad \delta\in \left[0,\frac{\pi}{2}\right)\cup\left(\frac{3\pi}{2},2\pi\right],\label{cos}
\end{equation}
where $p_0=(p_{0i}p_{0}^{i})^{1/2}$. Therefore, for any case one receives
\begin{eqnarray}
\tilde{c}&=&c\cos^{1-\cos\delta}\delta,\\
\tilde{m}&=&m\cos^{-2(1-\cos\delta)}\delta,\\
\mu&=&\cos^{-4(1-\cos\delta)}\delta-1,\\
\tilde{E}^2&=&m^2c^4+p^2c^2\cos^{2(1-\cos\delta)}\delta\\
&=&m^2c^4+p_0^2c^2\cos^{-2\cos\delta}\delta,\\
\mathcal{E}^2&=&-p^2c^2\left(1-\cos^{2(1-\cos\delta)}\delta\right)\\
&=&p_0^2c^2\left(\cos^{-2\cos\delta)}\delta-\cos^{-2}\delta \right),\\
\tilde{v}&=&v\left(\frac{m^2c^4+p^2c^2}{m^2c^4+p^2c^2\cos^{2(1-\cos\delta)}\delta}\right)^{1/2}\cos^{2(1-\cos\delta)}\delta\\
&=&v\left(\frac{p_0^2c^2+m^2c^4\cos^2\delta}{p_0^2c^2+m^2c^4\cos^{2\cos\delta}\delta}\right)^{1/2}\cos^{1-\cos\delta}\delta,\label{vela}
\end{eqnarray}
Consequently, for a massive particle
\begin{eqnarray}
\tilde{\Gamma}&=&\left(1-\frac{\tilde{v}^2}{c^2}\cos^{-2(1-\cos\delta)}\delta\right)^{-1/2},\\
\tilde{E}&=&mc^2\left(1-\frac{\tilde{v}^2}{c^2}\cos^{-2(1-\cos\delta)}\delta\right)^{-1/2},\label{conny}\\
\mathcal{E}^2&=&m^2 c^4\frac{1-\cos^{2(1-\cos\delta)}\delta}{\cos^{2(1-\cos\delta)}\delta}\frac{\tilde{v}^2}{c^2}\left(\frac{\tilde{v}^2}{c^2}-\cos^{2(1-\cos\delta)}\delta\right)^{-1},\\
p&=&m\tilde{v}\left(1-\frac{\tilde{v}^2}{c^2}\cos^{-2(1-\cos\delta)}\delta\right)^{-1/2}\cos^{-2(1-\cos\delta)}\delta,\\
\tilde{L}&=&-mc^2\left(1-\frac{\tilde{v}^2}{c^2}\cos^{2\cos\delta-1}\delta \right)\left(1-\frac{\tilde{v}^2}{c^2}\cos^{-2(1-\cos\delta)}\delta\right)^{-1/2},
\end{eqnarray}
and since the momentum magnitude is given through Special Relativity
\begin{equation}
p=mv\left(1-\frac{v^2}{c^2}\right)^{-1/2},\label{momo}
\end{equation}
one obtains
\begin{eqnarray}
\tilde{v}&=&v\left(1-\frac{v^2}{c^2}\left(1-\cos^{2(1-\cos\delta)}\delta\right)\right)^{-1/2}\cos^{2(1-\cos\delta)}\delta,\label{ve1}\\
\mathcal{E}^2&=&-m^2c^4\left(1-\cos^{2(1-\cos\delta)}\delta\right)\frac{v^2}{c^2}\left(1-\frac{v^2}{c^2}\right)^{-1},\\
\tilde{\Gamma}&=&\left(1-\frac{v^2}{c^2}\right)^{-1/2}\left(1-\frac{v^2}{c^2}\left(1-\cos^{2(1-\cos\delta)}\delta\right)\right)^{1/2},\\
\tilde{E}&=&mc^2\left(1-\frac{v^2}{c^2}\right)^{-1/2}\left(1-\frac{v^2}{c^2}\left(1-\cos^{2(1-\cos\delta)}\delta\right)\right)^{1/2},\label{connyA}\\
\tilde{L}&=&-mc^2\left(1-\frac{v^2}{c^2}\right)^{1/2}\left(1-\frac{v^2}{c^2}\left(1-\cos^{2(1-\cos\delta)}\delta\right)\right)^{-1/2}.
\end{eqnarray}
Furthermore, the consistency of the deformed speed formula (\ref{ve1}) demands existence of the maximal speed which in general differs from the speed of light in vacuum and depends on angular deformation
\begin{equation}
v<v_{max},\quad v_{max}=c\left(1-\cos^{2(1-\cos\delta)}\delta\right)^{1/2}\leq c,
\end{equation}
and always $\tilde{v}\leq v$ , $\tilde{v}\leq\tilde{c}$ and $v< c$ for a real and positive mass, while the relation (\ref{cos}) joined with (\ref{momo}) written for initial data gives the Eq. (\ref{eini}) and
\begin{equation}
\cos\delta=\frac{p_0}{mv}\left(1-\frac{v^2}{c^2}\right)^{1/2}=\frac{v_0}{v}\left(\frac{c^2-v^2}{c^2-v_0^2}\right)^{1/2},\quad v_0=\frac{p_0c^2}{E_0},
\end{equation}
and, for this reason, one obtains the applicability conditions for the theory
\begin{equation}
v_0\leq v\leq v_{max},\quad \ln\left|1+\frac{p_0^2}{m^2c^2}\right|\leq -2(1-\cos\delta)\ln|\cos\delta|.
\end{equation}

Similarly, according to Special Relativity, for a massless particle $m^2=0$
\begin{equation}
p=\frac{E}{v},\qquad v=c,
\end{equation}
and, for this reason, in the deformed theory for such particle
\begin{eqnarray}
\tilde{v}&=&v\cos^{1-\cos\delta}\delta\leq v,\label{mom}\\
\tilde{E}&=&p_0c\cos^{-\cos\delta}\delta=pc\cos^{1-\cos\delta}\delta,\label{fora1}\\
\mathcal{E}^2&=&p_0^2c^2\left(\cos^{-2\cos\delta}\delta-\cos^{-2}\delta\right)=-p^2c^2\left(1-\cos^{2(1-\cos\delta)}\delta\right),\\
\tilde{L}&=&-p_0c(1-\cos\delta)\cos^{-\cos\delta}\delta=-pc(1-\cos\delta)\cos^{1-\cos\delta}\delta,\label{fora3}
\end{eqnarray}
and, moreover, one receives
\begin{equation}
\cos\delta=\frac{E_0}{E},\qquad E_0=p_0c.
\end{equation}

For an imaginary mass $m=i|m|$, which in Special Relativity describes a tachyon, one has
\begin{equation}
p=|m|v\left(\frac{v^2}{c^2}-1\right)^{-1/2},\qquad v>c, \qquad E_0^2=p_0^2c^2-|m|^2c^4,
\end{equation}
and, therefore the speed formula (\ref{ve1}) remains unchanged, while
\begin{eqnarray}
\mathcal{E}^2&=&-|m|^2c^4\left(1-\cos^{2(1-\cos\delta)}\delta\right)\frac{v^2}{c^2}\left(\frac{v^2}{c^2}-1\right)^{-1},\\
\tilde{\Gamma}&=&\left(\frac{v^2}{c^2}-1\right)^{-1/2}\left(1-\frac{v^2}{c^2}\left(1-\cos^{2(1-\cos\delta)}\delta\right)\right)^{1/2},\\
\tilde{E}&=&|m|c^2\left(\frac{v^2}{c^2}-1\right)^{-1/2}\left(1-\frac{v^2}{c^2}\left(1-\cos^{2(1-\cos\delta)}\delta\right)\right)^{1/2},\label{connyA}\\
\tilde{L}&=&-|m|c^2\left(\frac{v^2}{c^2}-1\right)^{1/2}\left(1-\frac{v^2}{c^2}\left(1-\cos^{2(1-\cos\delta)}\delta\right)\right)^{-1/2},
\end{eqnarray}
and, moreover,
\begin{equation}
\cos\delta=\frac{p_0}{|m|v}\left(\frac{v^2}{c^2}-1\right)^{1/2}=\frac{v_0}{v}\left(\frac{v^2-c^2}{v_0^2-c^2}\right)^{1/2},\quad v_0=\frac{p_0c^2}{E_0}>c,
\end{equation}
with the consistency condition
\begin{equation}
v_{max}\leq v_0\leq v,\quad \ln\left|1-\frac{p_0^2}{|m|^2c^2}\right|\geq -2(1-\cos\delta)\ln|\cos\delta|.
\end{equation}
If $v>c$, then also $\tilde{v}>\tilde{c}$ and $\tilde{v}>c$, and the deformed speed of a tachyon written explicitly through the initial value $p_0$
\begin{equation}
\tilde{v}=v\left(\frac{p_0^2c^2-|m|^2c^4\cos^2\delta}{p_0^2c^2-|m|^2c^4\cos^{2\cos\delta}\delta}\right)^{1/2}\cos^{1-\cos\delta}\delta,
\end{equation}
shows that a tachyon exists for $\cos\delta\in (0,1)$, that is for the angles
\begin{equation}
\delta\in \left(0,\frac{\pi}{2}\right)\cup\left(\frac{3\pi}{2},2\pi\right).
\end{equation}
In other words, when angular deformation is non-zero, then tachyon's existence has a more physical justification compared to its merely formal nature in Special Relativity. Whenever $\delta^2<0$, then one should make a replacement $\delta\rightarrow i|\delta|$ or $\cos\delta\rightarrow\cosh|\delta|$ in all formulas with non-trivial deformation, and, therefore, an imaginary deformation gives a hyperbolic nature to the theory.

\section{Angularly Deformed Quantum Mechanics}

\subsection{Relativistic Theory}
A contravariant deformed energy-momentum four-vector is $\tilde{p}^\mu =[\tilde{E},p^i\tilde{c}]$, and a covariant one is $\tilde{p}_\mu=\eta_{\mu\nu}\tilde{p}^\nu$. Writing the constraint (\ref{bomb1}) as follows
\begin{equation}
\tilde{p}_\mu \tilde{p}^\mu-\tilde{m}^2\tilde{c}^4=0,
\end{equation}
one can apply the deformed relativistic canonical quantization
\begin{equation}
\tilde{p}^\mu\rightarrow \hat{\tilde{p}}^\mu=i\tilde{\hbar} \tilde{c}\tilde{\partial}^\mu,
\end{equation}
where for curvilinear space-time coordinates $\tilde{x}^\mu=[\tilde{c}\tilde{t},\tilde{x}^i]=\tilde{x}^\mu(x^\mu)$ one has
\begin{equation}
\tilde{\partial}^\mu=\eta^{\mu\nu}\tilde{\partial}_\nu, \quad \tilde{\partial}_\mu=\frac{\partial}{\partial \tilde{x}^\mu}=\left[\tilde{\partial}_{0},\tilde{\partial}_i\right],\quad \tilde{\partial}_{0}=\frac{1}{\tilde{c}}\frac{\partial}{\partial\tilde{t}},\quad \tilde{\partial}_i=\frac{\partial}{\partial \tilde{x}^i},
\end{equation}
and one receives the deformed relativistic quantum mechanics
\begin{eqnarray}
\left(\eta_{\mu\nu}\tilde{\partial}^\mu\tilde{\partial}^\nu+\frac{\tilde{m}^2\tilde{c}^2}{\tilde{\hbar}^2}\right)\tilde{\Psi}(\tilde{x},\tilde{t})&=&0,\label{dKG1}\\
\left(i\tilde{\hbar}\tilde{c}\gamma^\mu\tilde{\partial}_\mu-\tilde{m}\tilde{c}^2\right)\tilde{\psi}(\tilde{x},\tilde{t})&=&0,\label{dDir1}
\end{eqnarray}
where $\gamma^\mu$ are the Dirac gamma matrices. For the coordinate transformation given through the local dilatation transformations
\begin{equation}
  d\tilde{x}^\mu=\Omega dx^\mu\quad,\quad dx^\mu=\Omega^{-1}d\tilde{x}^\mu,\label{dilat}
\end{equation}
the deformed theories (\ref{dKG1}) and (\ref{dDir1}) transform into the standard theories if and only if the scale function has the form
\begin{equation}
\Omega=\frac{\tilde{\hbar}\tilde{c}}{\hbar c}.
\end{equation}
Thus, for these specific local conformal transformations of space-time coordinates, there arises the local emergent deformed space characterized through the conformal metric
\begin{equation}
  \tilde{\eta}_{\mu\nu}=\eta_{\alpha\beta}\frac{\partial \tilde{x}^\alpha}{\partial x^\mu}\frac{\partial \tilde{x}^\beta}{\partial x^\nu}=\Omega^2\eta_{\mu\nu},\quad
  \tilde{\eta}^{\mu\nu}=\Omega^{-2}\eta^{\mu\nu},\quad \tilde{\eta}^{\mu\alpha}\tilde{\eta}_{\alpha\nu}=\delta^\mu_\nu,
\end{equation}
where $\delta^\mu_\nu=\mathrm{diag}[1,1,1,1]$, which allows to establish the derivatives
\begin{equation}
\tilde{\partial}_\mu=\Omega^{-1}\partial_\mu\quad,\quad \tilde{\partial}^\mu=\Omega\tilde{\eta}^{\mu\nu}\partial_\nu=\Omega^{-1}\partial^\mu,
\end{equation}
and shows that relativistic invariance on the emergent deformed space-time is preserved locally and globally. For considerations of quantum theory one can reduce the analysis to the local level, wherein the dilatation transformations of space-time coordinates
\begin{equation}
\tilde{x}'^\mu=\Omega\left(\tilde{\Lambda}^\mu_{~\!\nu} x^\nu+x_0^\mu\right),
\end{equation}
assure relativistic invariance whenever
\begin{equation}
d\tilde{x}'_\mu=\Omega\tilde{\Lambda}^\nu_{~\!\mu} dx_\nu,\quad \Omega=\Omega_0,
\end{equation}
where $\Omega_0$ is a constant, and in the moving-particle representation
\begin{equation}
\tilde{\Lambda}^\mu_{~~\!\nu}=\left[\begin{array}{cc}\tilde{\Gamma}&-\tilde{\Gamma} R^i_j\frac{\tilde{v}^i}{\tilde{c}}\\
-\tilde{\Gamma}\frac{\tilde{v}_j}{\tilde{c}}&R^i_k\left(\delta^k_j+(\tilde{\Gamma}-1)\frac{\tilde{v}^k\tilde{v}_j}{\tilde{v}^2}\right)\end{array}\right],\qquad\tilde{\Gamma}=\left(1-\frac{\tilde{v}^2}{\tilde{c}^2}\right)^{-1/2}.
\end{equation}
Then, the emergent deformed space-time is flat, homogenous, and isotropic, and belongs to a wide class of the Robertson--Walker cosmological metrics.

The Lie algebra $\widetilde{\mathfrak{iso}}(1,3)=\widetilde{\mathfrak{so}}(1,3)\dot{+}\tilde{\mathfrak{r}}$ of the deformed Poincar\'{e} group is given through the semisimple deformed Lorentz algebra
\begin{equation}
\widetilde{\mathfrak{so}}(1,3)=\mathrm{span}\left\{\tilde{M}^{\lambda\mu}\right\},
\end{equation}
and the Abelian radical
\begin{equation}
  \tilde{\mathfrak{r}}=\mathrm{span}\left\{\tilde{P}^\mu\right\},
\end{equation}
where the deformed operators of momentum and angular momentum are
\begin{eqnarray}
    \tilde{P}^\mu&=&i\tilde{\hbar} \tilde{c}\tilde{\partial}^\mu=P^\mu,\label{op1}\\
    \tilde{M}^{\lambda\mu}&=&i\tilde{\hbar} \tilde{c}\left(\tilde{x}^\lambda\tilde{\partial}^\mu-\tilde{x}^\mu\tilde{\partial}^\lambda\right)=\Omega_0 M^{\lambda\mu},\label{op2}
\end{eqnarray}
with $P^\mu$ and $M^{\lambda\mu}$ being the operators defining the non-deformed algebra, and forms the deformed Lie algebra
\begin{eqnarray}
 \left[\tilde{x}^\lambda,\tilde{P}^\mu\right]&=&-i\tilde{\hbar}\tilde{c}\tilde{\eta}^{\lambda\mu},\\
  \left[\tilde{P}^\lambda,\tilde{P}^\mu\right]&=&0,\\
  \left[\tilde{M}^{\lambda\mu},\tilde{P}^\nu\right]&=&i\tilde{\hbar} \tilde{c}\left(\tilde{\eta}^{\mu\nu}\tilde{P}^\lambda-\tilde{\eta}^{\lambda\mu}\tilde{P}^\mu\right),\\
  \left[\tilde{M}^{\lambda\mu},\tilde{M}^{\rho\sigma}\right]&=&-i\tilde{\hbar} \tilde{c}\left(\tilde{\eta}^{\lambda\rho}\tilde{M}^{\mu\sigma}-\tilde{\eta}^{\mu\rho}\tilde{M}^{\lambda\sigma}+\tilde{\eta}^{\mu\sigma}\tilde{M}^{\lambda\rho}-\tilde{\eta}^{\lambda\sigma}\tilde{M}^{\mu\rho}\right),
\end{eqnarray}
which written explicitly is apparently inequivalent to the non-deformed algebra throughout presence of the scale factor
\begin{eqnarray}
 \left[x^\lambda, P^\mu\right]&=&-i\frac{{\hbar}{c}}{\Omega_0^{2}}{\eta}^{\lambda\mu},\\
  \left[P^\lambda,{P}^\mu\right]&=&0,\\
\left[{M}^{\lambda\mu},P^\nu\right]&=&i\frac{{\hbar}{c}}{\Omega_0^{2}}\left({\eta}^{\mu\nu}{P}^\lambda-{\eta}^{\lambda\mu}{P}^\mu\right),\\
  \left[{M}^{\lambda\mu},{M}^{\rho\sigma}\right]&=&-i\frac{{\hbar}{c}}{\Omega_0^{2}}\left({\eta}^{\lambda\rho}{M}^{\mu\sigma}-{\eta}^{\mu\rho}{M}^{\lambda\sigma}+{\eta}^{\mu\sigma}{M}^{\lambda\rho}-{\eta}^{\lambda\sigma}{M}^{\mu\rho}\right).
\end{eqnarray}
For this Lie algebra there are two independent Casimir operators, the first one is squared deformed momentum operator
\begin{equation}
  \tilde{P}^2=\tilde{P}_\mu \tilde{P}^\mu=\eta_{\mu\nu}\tilde{P}^\mu\tilde{P}^\nu=\tilde{m}^2\tilde{c}^4=m^2c^4=P^2,
\end{equation}
whereas the second one is square of the deformed Pauli-Lubanski four-vector
\begin{equation}
  \tilde{W}_{\lambda}=\frac{1}{2}\varepsilon_{\lambda\mu\nu\rho}\tilde{P}^\mu \tilde{M}^{\nu\rho},
\end{equation}
which can be easily derived
\begin{equation}
\tilde{W}^2=\eta_{\mu\nu}\tilde{W}^\mu\tilde{W}^\nu=-(\tilde{\hbar}\tilde{c})^2m^2c^4 s(s+1),\label{spin}
\end{equation}
where $s$ is a spin eigenvalue, having the following properties
\begin{eqnarray}
\tilde{P}_\mu\tilde{W}^\mu&=&0,\\
\left[\tilde{W}^{\lambda},\tilde{P}^\mu\right]&=&0,\\
\left[\tilde{W}^{\lambda},\tilde{M}^{\mu\lambda}\right]&=&i\tilde{\hbar}\tilde{c}\left(\tilde{\eta}^{\lambda\mu}\tilde{W}^\nu-\tilde{\eta}^{\lambda\nu}\tilde{W}^\mu\right),\\
\left[\tilde{W}^{\lambda},\tilde{W}^{\mu}\right]&=&i\tilde{\hbar}\tilde{c}\varepsilon^{\lambda\mu\nu\rho}\tilde{W}_{\nu}\tilde{P}_\rho.
\end{eqnarray}
Actually, the Casimir operator (\ref{spin}) can be rewritten as follows
\begin{equation}
\tilde{W}^2=-(\hbar c)^2m^2c^4 \tilde{s}(\tilde{s}+1),\label{spin1}
\end{equation}
to determine the deformed spin eigenvalues
\begin{equation}\label{spinn}
\tilde{s}_{\pm}=-\frac{1}{2}\pm\left(\frac{1}{4}+\Omega_0^2s(s+1)\right)^{1/2}.
\end{equation}
Furthermore, for solutions of both the Dirac equation and Klein-Gordon equation, the local deformed Lorentz matrices
\begin{equation}
\tilde{\Lambda}^\mu_{~\!\nu}=\exp\left\{\frac{1}{2}\left(\tilde{\omega}^{\kappa\lambda}\right)^\mu_{~\!\nu}\theta_{\kappa\lambda}\right\}
\end{equation}
where the operators of infinitesimal Lorentz rotations
\begin{equation}
\left(\tilde{\omega}^{\kappa\lambda}\right)^\mu_{~\!\nu}=\eta^{\lambda\mu}\delta^\kappa_\nu-\eta^{\kappa\mu}\delta^\lambda_\nu,
\end{equation}
can be approximated
\begin{equation}
  \tilde{\Lambda}^\mu_{~\!\nu}=\delta^\mu_{~\!\nu}+\frac{1}{2}\left(\tilde{\omega}^{\kappa\lambda}\right)^\mu_{~\!\nu}\theta_{\kappa\lambda}\quad,\quad (\tilde{\Lambda}^{-1})^\mu_{~\!\nu}=\delta^\mu_{~\!\nu}-v\left(\tilde{\omega}^{\kappa\lambda}\right)^\mu_{~\!\nu}\theta_{\kappa\lambda},
\end{equation}
and the relativistic invariance for solutions gives the infinitesimal generator as
\begin{equation}
  \tilde{J}^{\mu\nu}=\frac{i\tilde{\hbar}\tilde{c}}{4}[\gamma^\mu,\gamma^\nu]+\tilde{M}^{\mu\nu}.\label{kopa}
\end{equation}
Remarkably, the Eq. (\ref{op2}) allows to write
\begin{equation}
  \tilde{J}^{\mu\nu}=\Omega_0J^{\mu\nu},\label{kopa1}
\end{equation}
where $J^{\mu\nu}$ is the infinitesimal generator of the Lorentz group of Special Relativity. From the formal point of view, the aforementioned Casimir elements of the Lie algebra $\widetilde{\mathfrak{iso}}(1,3)$ allow to construct the irreducible representations of this algebra, particularly infinite-dimensional unitary representations.

\subsection{Semi-Classical Non-Relativistic Theory}
In the semi-classical approximation, one has
\begin{equation}
\tilde{\mathrm{E}}=\frac{p^2}{2\tilde{m}}+\mathrm{\tilde{V}}(\tilde{x}),
\end{equation}
and quantization lead to the semi-classical non-relativistic quantum mechanics
\begin{equation}
 i\tilde{\hbar}\frac{\partial\tilde{\Phi}(\tilde{x},\tilde{t})}{\partial \tilde{t}}=\left(-\frac{\tilde{\hbar}^2}{2\tilde{m}}\tilde{\nabla}^2+\tilde{\mathrm{V}}(\tilde{x})\right)\tilde{\Phi}(\tilde{x},\tilde{t}),\label{schr}
\end{equation}
where $\tilde{\nabla}^2=\delta^{ij}\tilde{\partial}_i\tilde{\partial}_j$ is the Laplace operator, and the wave functions obey the following deformed normalization condition
\begin{equation}
\left(\int d^3\tilde{x}(|\det \tilde{\eta}_{ij}|)^{1/2}\right)^{-1}\int d^3\tilde{x}(|\det \tilde{\eta}_{ij}|)^{1/2}|\tilde{\Phi}(\tilde{x},\tilde{t})|^2=1.
\end{equation}
Under the aforementioned local dilatations, the emergent deformed space-time is determined through the following curvilinear coordinates
\begin{equation}
\tilde{t}=\frac{c}{\tilde{c}}\Omega_0t,\quad\tilde{x}^i=\Omega_0x^i,
\end{equation}
and the theory (\ref{schr}) takes the form
\begin{equation}
 i\hbar\frac{\partial\tilde{\Phi}(x,t)}{\partial t}=\left(-\frac{\hbar^2}{2m}\nabla^2+\tilde{\mathrm{V}}(x)\right)\tilde{\Phi}(x,t),\label{schrW}
\end{equation}
that is if one identifies $\tilde{\Phi}(x,t)$ with the Schr\"odinger wave function and $\tilde{\mathrm{V}}(x)$ with a standard potential, then one receives the Schr\"odinger equation. Meanwhile, in the deformed space-time the theory (\ref{schr}) becomes
\begin{equation}
 i\hbar\frac{\partial\tilde{\Phi}(x,t)}{\partial t}=\left(-\frac{\tilde{\hbar}^2}{2\tilde{m}}\tilde{\eta}^{ij}\partial_i\partial_j+\tilde{\mathrm{V}}(x)\right)\tilde{\Phi}(x,t),\label{schrWa}
\end{equation}
and, for this reason, from the phenomenological point of view the presence of the scale factor opens the possibility for the new physics related to the emergent deformed space-time. If one considers the wave functions
\begin{equation}
\tilde{\Phi}(x,t)=\tilde{R}(x)\exp\left(-\frac{i}{\hbar}\tilde{\mathrm{E}}t\right),
\end{equation}
then in the deformed space-time one has the stationary quantum mechanics
\begin{equation}
\tilde{\mathrm{E}}\tilde{R}(x)=\left(-\frac{\tilde{\hbar}^2}{2\tilde{m}}\nabla^2+\mathrm{V}(x)\right)\tilde{R}(x).\label{schr1a}
\end{equation}
Furthermore, one can verify immediately that the following comutators hold
\begin{equation}
\left[\hat{\tilde{x}}_i,\hat{\tilde{p}}_j\right]=i\tilde{\hbar}\delta_{ij}\quad,\quad
\left[\hat{\tilde{x}}^i,\hat{\tilde{p}}_i\right]=i\tilde{\hbar}\cos\delta\quad,\quad
\left[\tilde{t},\hat{\tilde{\mathrm{E}}}\right]=i\tilde{\hbar}.
\end{equation}
Considering a quantum average in the deformed space
\begin{eqnarray}
\langle A\rangle=\left(\int d^3\tilde{x}(|\det \tilde{\eta}_{ij}|)^{1/2}\right)^{-1}\int d^3\tilde{x}\left(\det\tilde{\eta}_{ij}\right)^{1/2}\tilde{R}(\tilde{x})\hat{\tilde{A}}\tilde{R}(\tilde{x}),
\end{eqnarray}
where $\hat{\tilde{A}}=\hat{\tilde{A}}(\hat{\tilde{p}},\hat{\tilde{x}})$ is a quantum-mechanical operator, one can determine a standard deviation $\Delta A=\left(\left\langle(A-\langle A\rangle)^2\right\rangle\right)^{1/2}$, and, in particular, establish the deformed uncertainty principle
\begin{equation}
\Delta p_i\Delta \tilde{x}^i\geq\frac{\tilde{\hbar}}{2}\cos\delta,\quad \Delta \tilde{\mathrm{E}}\Delta \tilde{t}\geq\frac{\tilde{\hbar}}{2}.
\end{equation}

\section{Summary}

In this paper, we have considered an impact of the hypothetical violation of the velocity-momentum parallelism onto the Einstein equivalence principle, which led to a certain new insight into quantum mechanics. Although the deformed theory in itself has been constructed in the way of a straightforwardly analogy with Special Relativity, just the foundational assumption on existence of the angle $\delta$ gives rise to make revision of foundations of physics throughout experimental and observational data. From a technical point of view, the obtained model of deformed Special Relativity has involved the technique of an eikonal equation, which in itself is one of the pillars for modern optics and quantum mechanics, and by virtue of this method the deformed theory has been able to generate a new physics. From the phenomenological point of view, the most intriguing aspect of the deformed theory is the ability of this model to meet a positive empirical verification. In particular, the problem of tachyons as well as relativity-based measurement, for which non-zero value of $\delta$ can be related to an observational error.

Furthermore, the deformed theory displays a remarkable symmetry with respect to the action of the deformed Poincar\'{e} transformations, wherein the deformed Lorentz transformations form a subgroup, what from the phenomenological point of view establishes theoretical correctness of new derivations. This aspect has been generated the essentially new ideas on relativistic quantum mechanics, which are able to elucidate the nature of bosons and fermions. Furthermore, the main result following existence of a non-zero $\delta$ is emergence of the local deformed space-time, which is given through the established dilatation curvilinear coordinates and whose metric displays the features of a cosmological space-time. Another important result of the presented theory is the new insight into the nature of the Schr\"odinger equation, which under some assumptions becomes invariant with respect to the action of the deformation and, for this reason, can be regarded as the window which joins a non-deformed and deformed space-time. Remarkably, although the foundational symmetries of Special Relativity are basically preserved when the deformation is present, just this presence offers new possibilities in the matter of both theory and phenomenology of high energy physics on the basis of the further ramificaions from the deformed relativistic quantum mechanics to quantum field theory.

It should be emphasized that actual phenomenological applicability of the idea of deformation caused through $\delta$ can be efficiently verified throughout a direct conffrontation with the experimental data of particle physics and observational data of ultra-high energy astrophysics. Particularly, with help of the presented theory, the results manifestly beyond the Standard Model could be efficiently explained, whereas some aspects of the phenomena such like dark energy or dark matter can be also consistently described throughout presence of the angular deformation, because just the constraint (\ref{constra}) can be interpreted as existence of an extra energy caused through the deformation of the Einstein equivalence between mass, momentum, and energy of a moving particle. From the point of view of the theory of measurement, the deformed quantities rather than non-deformed ones should be the object of a measurement and, threfeore, be the physical quantities, because looking from the point of view of relativistic symmetries, just the deformed quantities are invariants of the Lorentz transformations. Regarding the nature of the angle $\delta$, since according to the analysis its value can be measured through measurement of invariant quantities, the angular deformation in itself is a relativistically invariant quantity. Discussion of the proposed theory in the context of the presently available observational constraints, which were not taken into account in this paper through its purely theoretical character, is worthy of further development.

\end{document}